%% file: main.tex
\documentclass[sigconf, screen]{acmart}

\usepackage[inkscapeformat=png]{svg}
\usepackage{svg}

\input{mymacros}

\copyrightyear{2023} 
\acmYear{2023} 
\setcopyright{rightsretained} 
\acmConference[ACE '24]{}{}{}
\acmBooktitle{ACE '24}
\acmDOI{}
\acmISBN{}

\title[Next-Step Hint Generation for Introductory Programming Using Large Language Models]{Next-Step Hint Generation for Introductory Programming\\ Using Large Language Models}   

\author{Lianne Roest}      
\affiliation{
  \institution{Utrecht University}
  \country{Utrecht, The Netherlands}
}
\email{lianneroest@gmail.com}

\author{Hieke Keuning}
\affiliation{
  \institution{Utrecht University}
  \country{Utrecht, The Netherlands}
}
\email{h.w.keuning@uu.nl}

\author{Johan Jeuring}
\affiliation{
  \institution{Utrecht University}
  \country{Utrecht, The Netherlands}
}
\email{j.t.jeuring@uu.nl}

\begin{document}

\ccsdesc[500]{Social and professional topics~Computer science education}

\keywords{Next-step hints, automated feedback, learning programming, Large Language Models, Generative AI}

\include{abstract}  

\maketitle                  

\input{Chapters/Chapter1/1.1introduction}
\input{Chapters/Chapter1/1.2researchquestions}

\input{Chapters/Chapter2/Programminghints}

\input{Chapters/Chapter3/LLMs}

\input{Chapters/Chapter4/4.1method}

\input{Chapters/Chapter4/4.2evaluation}

\input{Chapters/Chapter5/Promptengineering}

\input{Chapters/Chapter6/Genertingnextstephints}

\input{Chapters/Chapter7/Discussion}
\input{Chapters/Chapter7/Conclusion}


\bibliographystyle{ACM-Reference-Format}
\bibliography{refs}

\end{document}

%% file: mymacros.tex
\usepackage{float}
\usepackage{graphicx}
\usepackage{amsmath}
\usepackage{tabularx}
\usepackage{hyperref}
\usepackage{xspace}
\usepackage{dirtytalk}
\usepackage{listings}

\hypersetup{
    colorlinks,
    citecolor=black,
    filecolor=black,
    linkcolor=black,
    urlcolor=black
}

\newcommand{\eg}{\emph{e.g.,}\xspace}

\newcommand{\etal}{et~al.\xspace}

\newcommand{\ttt}{\textit{TaskTracker-tool}}

\renewenvironment{quote}{%
  \list{}{%
    \leftmargin0.5cm   
    \rightmargin\leftmargin
  }
  \item\relax
}
{\endlist}

%% file: abstract.tex
\begin{abstract}
Large Language Models possess skills such as answering questions, writing essays or solving programming exercises. 
Since these models are easily accessible, researchers have investigated their capabilities and risks for programming education. This work explores how LLMs can contribute to programming education by supporting students with automated next-step hints.
We investigate prompt practices that lead to effective next-step hints and use these insights to build our StAP-tutor. We evaluate this tutor by conducting an experiment with students, and performing expert assessments. Our findings show that most LLM-generated feedback messages describe one specific next step and are personalised to the student's code and approach. However, the hints may contain misleading information and lack sufficient detail when students approach the end of the assignment. This work demonstrates the potential for LLM-generated feedback, but further research is required to explore its practical implementation.    
\end{abstract}

%% file: Chapters/Chapter1/1.1introduction.tex
\section{Introduction}

Over the last decades, researchers have been developing a variety of digital tools and systems that support students in learning how to program. Most of these tools provide automated feedback on student solutions to exercises 
\cite{keuning2018systematic}. This feedback may include hints for code edits, references to relevant concepts, or suggestions for reading material. In this paper we will focus on \textit{next-step hints}. These hints help students with how to proceed when they are stuck while solving a programming exercise.

Next-step hints can be generated automatically by comparing possibly incomplete student programs with model solutions~\cite{keuning2014strategy}, or with historical data from other students~\cite{rivers2017data,price2019comparison}. 
Some issues with data-driven approaches are that they often require a large amount of data. Furthermore, these hints are often exact directions for what a student should do, revealing an answer and thus interfering with learning opportunities \cite{price2017factors}.
Adding explanations can improve students' understanding of the hints and perceived relevance \cite{marwan2019impact}.
However, handcrafting explanations is a lot of work, nullifying the advantage of data-driven approaches regarding development time.

In this paper we examine how large language models (LLMs) can be used to develop next-step hints. When we use LLMs to generate next-step hints, we don't need historical student data or model solutions anymore. The performance of LLMs has increased significantly in the last couple of years. LLMs are trained on enormous amounts of data and have shown to be very good at specific tasks, such as generating text, images, and code \cite{chen2021evaluating}. Especially with the arrival of the models from OpenAI (GPT-3, Codex, and GPT-4), LLMs have received much attention, both in and outside the academic world. Studies have shown that they are very good at solving introductory programming assignments \cite{denny2023conversing, finnie2022robots, prather2023robots}. As these models are now also available for students, there are concerns regarding plagiarism, integrity, and learning \cite{finnie2022robots, becker2023programming}.

Novice programmers may be particularly susceptible to the `dangers' of using LLMs as their pair programmers. They do not have the skills yet to understand code and recognise code of good quality, and might not know how to design effective prompts \cite{doderlein2022piloting}. Moreover, students might overestimate their coding abilities using LLMs, leading to over-reliance and reduced learning \cite{prather2023robots}.


%% file: Chapters/Chapter1/1.2researchquestions.tex

In this study, we focus on introductory Python exercises, as Python is a programming language often used in novice programming courses. We will investigate how we can design prompts for LLMs that produce next-step hints, and enhance them with additional information, such as explanations. We want to go beyond just edit instructions, as often generated with data-driven methods. Based on our findings, we built the StAP-tutor (\textbf{St}ep \textbf{A}ssisted \textbf{P}rogramming tutor). 
Finally, we will evaluate the quality of the generated hints with an experiment with students and an expert assessment. 
We investigate the following main research question and two subquestions:

\begin{enumerate}
    \item[\textbf{RQ1}] To what extent can we use LLMs to generate informative and effective next-step hints for Python introductory programming exercises?
    \item[\textbf{SQ1}] What prompt characteristics are suitable for generating effective next-step hints with LLMs?
    
    \item[\textbf{SQ2}] What are students' and experts' perceptions of the quality of LLM-generated next-step hints?
\end{enumerate}

Section~\ref{sec:relatedwork} discusses related work on effective feedback, generating automated feedback, and using LLMs in programming education. We present a short overview of the method in Section~\ref{sec:method}, with a more in-depth discussion on prompt-engineering in Section~\ref{chap:prompt}. Section~\ref{sec:evaluation} describes the tool and its evaluation. Section~\ref{sec:discussion} discusses limitations and directions for future research. Section~\ref{sec:conclusions} concludes.

%% file: Chapters/Chapter2/Programminghints.tex
\section{Related work}
\label{sec:relatedwork}

\subsection{Automated feedback}

Digital learning tools have been developed across multiple educational domains, with many teaching topics within computer science and programming \cite{mousavinasab2021intelligent}. 
An important aspect of these educational programming tools is providing (automated) feedback \cite{deeva2021review}. 

\subsubsection{Providing effective feedback}\label{lit: effb}
Feedback is considered essential for learning \cite{hattie2007power}. It can correct mistakes or misconceptions, and improve motivation through positive comments or guidance when the student is stuck. \textit{Formative} feedback consists of information or learning activities that support and facilitate student learning \cite{irons2021enhancing}. \textit{Summative} feedback consists of assessment activities, resulting in grades that evaluate a student's performance. 

Different variables impact the effectiveness of formative feedback  \cite{shute2008focus}. \textit{Elaborated} feedback helps students understand why something is wrong and can guide students in the right direction with explanations or tips. However, if feedback is too long, learners will not pay attention to it, rendering it useless. Another essential factor is \textit{timing}. Feedback can be provided by intervening or waiting until the student requests feedback themselves or finishes an assignment. Another variable of feedback is its \textit{nature}, which can be positive or negative. In contrast with negative feedback, positive feedback reinforces what students are doing well, for example, to emphasise a student is correctly implementing assignment requirements. Both can, however, have beneficial effects on learning \cite{hattie2007power}.

Defining effective feedback is not easy, and students and teachers have sometimes different ideas about it~\cite{dawson2019makes}. 
While teachers believe that timing, modalities and connected tasks are important, students prefer detailed feedback personalised to the student's work. 

\subsubsection{Feedback on programming tasks}\label{lit:auto:types}

Narciss \cite{narciss2008feedback} identified five main categories for the content of feedback: 
Knowledge About Task Constraints, Knowledge About Concepts, Knowledge About Mistakes, Knowledge About How to Proceed, and Knowledge About Meta-cognition.
For the domain of programming, Keuning \etal~\cite{keuning2018systematic} conducted a systematic review of automated programming feedback in which they extended Narciss' classification for this domain. 
They found that most programming learning tools provide feedback with knowledge about mistakes by implementing automated testing, as this approach is generally easy to implement. However, Hao \etal~\cite{hao2022towards} found that feedback of this type is not as effective as more informative and detailed feedback. Only presenting mistakes requires students to interpret the results themselves, which may interfere with learning from their misconceptions. 

This paper focuses on generating feedback on how to proceed.
These next-step hints can be of different forms, such as suggestions, questions or instructions. The hint's goal can be to correct an error or guide a student to take a next step towards a correct solution. 

Several techniques exist for generating automated programming feedback. Keuning \etal identified three universal techniques (model tracing, constraint-based modelling, and data analysis) and five domain-specific (e.g.~dynamic code analysis, and static analysis) to the field of programming.
The choice of feedback technique determines, to some extent, the type of feedback generated.

Classical approaches such as model tracing have been used for next-step hints, but they require a significant amount of manual labour \cite{anderson1985lisp,keuning2014strategy}. Every knowledge component, production rule, or constraint has to be defined by experts. Especially in the programming domain, where assignments often have different solutions, semantically or syntactically, this can be very time-consuming. 
Data analysis approaches overcome these issues by employing algorithms that use large existing datasets to learn patterns or strategies that lead to correct solutions from . 

\subsubsection{Data-driven next-step hints}\label{lit:data}

A substantial part of the data-driven approaches is related to next-step hints \cite{price2019comparison}. In general, a data-driven method for programming operates as follows: first, it compares the current student state (from the moment of the hint request) to a desired target state, often a correct solution. Then, an algorithm identifies the differences or required edits to transfer the current state into the target state. From there, the algorithm extracts one or more next steps \cite{malysheva2022algorithm, rivers2017data}. 

As data-driven methods avoid the need for many model solutions, they are considered less time-consuming than traditional techniques. Another advantage is they can generate hints for never-before-seen states \cite{rivers2017data, price2016generating, paassen2018continuous}. Furthermore, they are not language specific; they can be implemented for various programming languages without too many alterations. 

There are still some limitations to current data-driven approaches. First, their hints often only inform students what to do without further explanation or additional information. This may cause uncertainty or confusion, with a chance of students not following the given hints. Marwan \etal \cite{marwan2019impact} added textual explanations to next-step hints. They found that this feedback had a higher follow-up rate and interpretability score, and students had an improved perception of its relevance. Unfortunately, these explanations had to be handcrafted. Additionally, Rivers found that novices wanted more elaborate feedback compared to experienced students \cite{rivers2017data}.

Second, Price \etal~\cite{price2019comparison} identify situations where data-driven hints perform poorly overall, such as when students have code that diverges from obvious or standard solutions. Finally, even though data-driven approaches are considered less time-consuming, they often remain complex due to transformations and comparing student programs. These shortcomings illustrate the possibility of improving existing data-driven techniques.

%% file: Chapters/Chapter3/LLMs.tex
\subsection{Large language models}\label{lit:LLM}
Large language models are deep-learning models, trained on enormous amounts of data and can generate output given a specific task based on this data. A \textit{prompt} defines a request, question or instruction for what the user wants the model to generate. 
OpenAI's GPT-3, released in May 2020, is one of the breakthrough models \cite{brown2020language}.
Some models are specifically designed for code synthesis, such as Alpha-Code, 
codeBERT, 
and OpenAI's Codex. 
OpenAI's ChatGPT and GPT-4 have received a lot of attention. With its human-like conversations and easy accessibility, especially ChatGPT has been immensely popular.
GPT-4 outperforms other existing LLMs on several academic benchmarks \cite{OpenAI2023GPT4TR}. 


\subsubsection{LLM in programming education}
Since the arrival of LLMs, researchers have expressed their concerns about the negative impact these models might have on programming education \cite{prather2023robots}. Recent work shows that LLMs perform well on CS1/CS2 programming tasks \cite{finnie2022robots,finnie2023my,denny2023conversing,prather2023robots}, and novice programmers may be particularly susceptible to overreliance on LLMs. 
Furthermore, LLMs may produce syntactically incorrect code or contain constructs that are too complex and not appropriate for beginners \cite{denny2023computing}.

Using an LLM may also interfere with the problem-solving process. Students' focus may shift to how to get an LLM to do what they want instead of solving the assignment \cite{prather2023s}. Vaithilingam \etal~\cite{vaithilingam2022expectation} studied more experienced students using LLMs while coding and reported that they often had difficulties understanding, editing, and debugging code snippets created by Copilot. 
Kazemitbaar \etal~\cite{kazemitabaar2023studying} found that students with access to Codex performed slightly better overall than a group without access on a post-test. However, students with access made significantly more errors in coding tasks where they had to write code from scratch, possibly indicating an over-reliance. 

Several studies illustrate potential opportunities of LLMs to support students and teachers in programming education. LLMs have been used to generate programming assignments, clarify programming error messages, generate code explanations, and help solving bugs \cite{sarsa2022automatic, macneil2022generating, macneil2023experiences,zhang2022repairing,leinonen2023using}. 
 

There are a few studies on giving feedback using LLMs. 
Phung \etal~\cite{phung2023generating} used Codex to create feedback on syntax errors consisting of a fixed program with a corresponding explanation.
Kiesler \etal~\cite{kiesler2023exploring} analysed the feedback from LLMs on incorrect programming submissions, and found that some feedback could be useful, but misleading information was often present.
Hellas \etal~\cite{hellas2023exploring} collected code from students' help requests in an online programming course and used an LLM to give feedback. They found that LLMs often find at least one issue, but struggle to find all of them. Interestingly, although they asked the LLM not to include a model solution, it often did.
Our work complements these studies by focusing on generating next-step programming hints.


%% file: Chapters/Chapter4/4.1method.tex
\section{Method}
\label{sec:method}
\begin{figure*}[h!]
    \centering
    \includegraphics[width=0.9\textwidth]{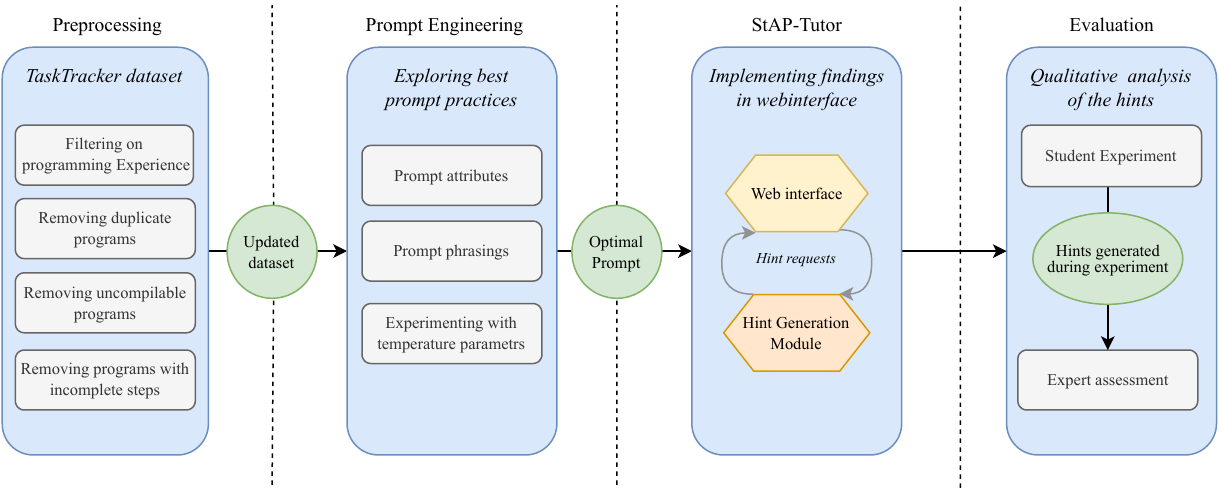}
    \caption{Method of this work.}
    \label{fig:workflow}
\end{figure*}

To answer our research questions we create a dataset with sequences of steps students take towards solving a programming problem (\ref{meth:dataset}), use these sequences to engineer a prompt for generating next-step hints (\ref{meth:prompteng}), build the StAP-tutor, and evaluate the results (\ref{meth:eval}). Figure \ref{fig:workflow} shows the workflow of this process. 

\subsection{Student program dataset}\label{meth:dataset}
To create a dataset of sequences of steps novice students take when solving a programming problem, we used the dataset from Lyulina \etal \cite{Lyulina2021tasktracker}, which contains snapshots of student programs for various programming tasks, collected with their \ttt. This tool saves a code snapshot with every keystroke. 
The dataset consists of programs from 148 participants with various ages and experience, and solutions in multiple languages. For this work, we only considered the Python solutions and programs from participants with less than half a year of programming experience. We selected two exercises that required applying the most operations and programming constructs: Pies and Brackets, shown in Table~\ref{table:exercisesprompt}.

\begin{table*}[h!]
\centering

\caption{Description of the exercises used for prompt engineering (pies and brackets) and the evaluation (clumps).}
\begin{tabular}{p{0.3\textwidth} p{0.3\textwidth} p{0.3\textwidth}}
\hline 
\textbf{Pies} & \textbf{Brackets} & \textbf{Clumps} \\ \hline 

A single pie costs A dollars and B cents in the cafe. Calculate how many dollars and cents one needs to pay for N pies. \newline \newline\textit{Input}: The program receives three numbers \newline A - how many dollars a pie costs;\newline B - how many cents a pie costs;\newline N - how many pies do you need to buy \newline \newline \textit{Output}: Print out two numbers: the cost of N pies in dollars and cents.
& Place opening and closing brackets into the input string like this: for odd length: example → e(x(a(m)p)l)e; for even length: card → c(ar)d, but not c(a()r)d.\newline \newline \textit{Input}: The program receives a string of English letters (lowercase and uppercase).\newline\newline \textit{Output}: Print out the string with the brackets added.   

&
Say that a \say{clump} in an array is a series of 2 or more adjacent elements of the same value. Return the number of clumps in the given array. For example, an array with the numbers {[}2,2,3,5,6,6,2{]} has 2 clumps. 
\newline\newline
\textit{Input}: The program receives a number n, followed by n lines with one integer per line. 
\newline\newline
\textit{Output}: Print out the number of clumps
\\

\hline
\end{tabular}
\label{table:exercisesprompt}
\end{table*}
We processed the data sequences as follows:

\begin{enumerate}
    \item \textit{Remove duplicate program states.}
    \item \textit{Remove program states with syntax errors.}
    We want to generate next-step hints, and not fix syntactically incorrect code. 
    \item \textit{Remove programs with incomplete steps.}
    We reduced the dataset by not considering single keystrokes; we only kept the last snapshot when a student was editing a line. We also removed snapshots in which students used print statements to trace their code and removed them shortly after, as they do not contribute to a program's functionality. 
    
\end{enumerate}



\subsection{Prompt engineering}\label{meth:prompteng}

To find out how we could best instruct the LLM (OpenAI's gpt-3.5-turbo model) to generate the desired output, 
we performed prompt engineering in an iterative process, where we determined how to proceed based on intermediate findings. We made calls to the OpenAI API with various prompts and settings to generate the hints.
Section \ref{chap:prompt} discusses this process in detail.

%% file: Chapters/Chapter4/4.2evaluation.tex
\subsection{Evaluation}\label{meth:eval}

We built and integrated our findings into the StAP-tutor. The StAP-tutor is a web interface where students can practice their Python skills with the help of next-step hints. 
As teachers and students have different perspectives on effective feedback, we performed two evaluations to assess the quality of the generated hints: an evaluation by experts, and an experiment with students. 

\subsubsection{Student evaluation} 

Before carrying out the experiment, we evaluated our method with the Ethics and Privacy Quick Scan of Utrecht University. They classified this research as low-risk, with no fuller ethics review or privacy assessment required.

\paragraph{Participants}
Based on convenience sampling, we recruited three first-year Bachelor of Artificial Intelligence students at Utrecht University in the Netherlands. They already had some programming experience and were taking a Python course. 

\paragraph{Exercise}
We let the students work on a different assignment than used for prompt engineering to validate if the final prompt generates hints of good quality for other assignments too. The exercise \say{clumpCount} is taken from the CodingBat.org website and shown in Figure \ref{table:exercisesprompt}. 
Its solution contains loops and conditionals, and is typically solved in multiple steps, making this exercise suitable for generating next-step hints. We adjusted the problem description to explicitly describe the input and output requirements, similar to the exercises used for prompt engineering. We found that adding these complete descriptions helped with references to variables from the problem description and student code in the hints. 

\begin{figure*}[h!]
    \centering
    \fcolorbox{lightgray}{white}{\includegraphics[width=0.7\textwidth]{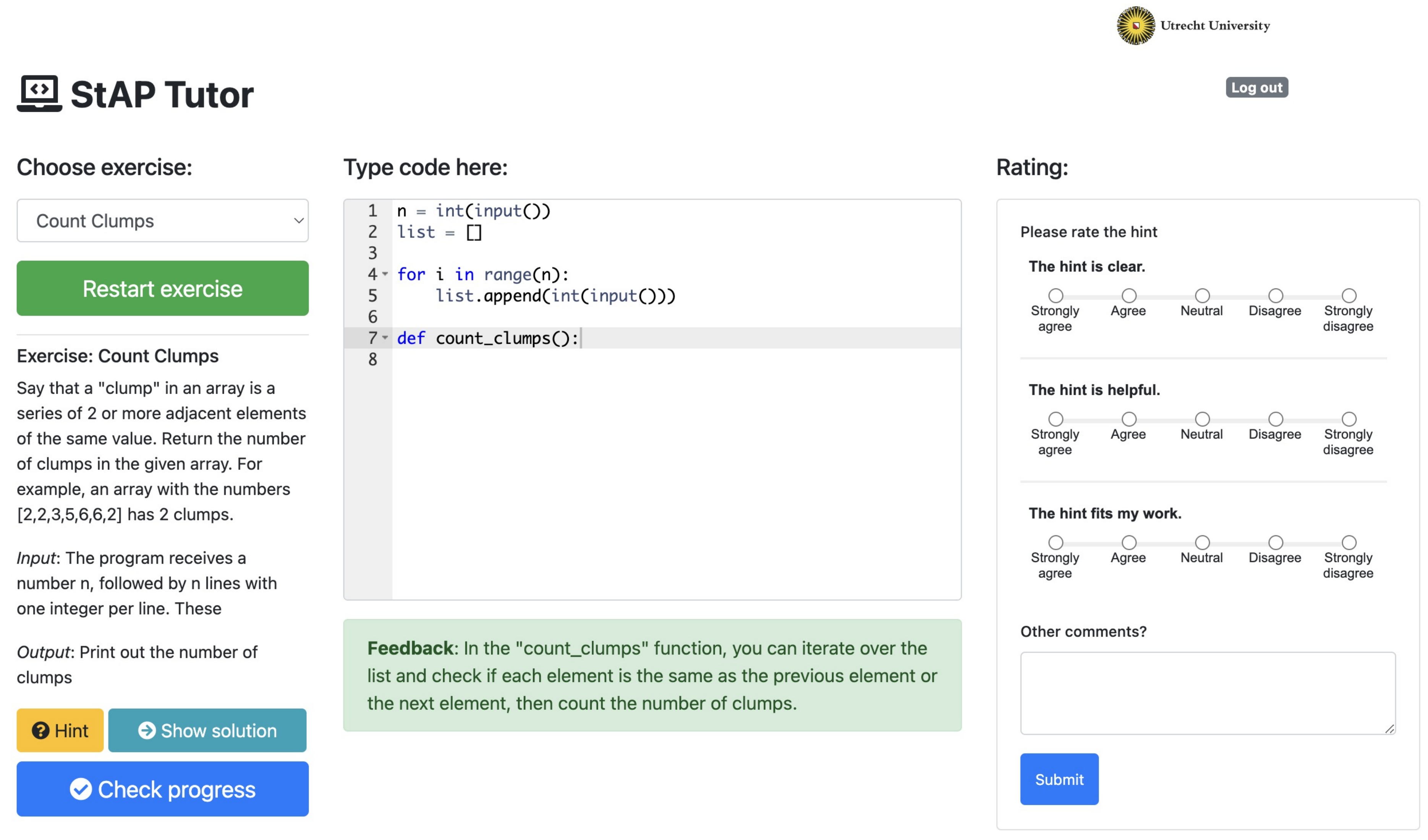}}
    \caption{Interaction with the StAP-tutor: the student is asked to rate the hint.}
    \label{fig:staptutorrate}
\end{figure*}

\paragraph{Experimental setup}

The duration of the experiment was roughly one hour.
After introducing the experiment, we assigned each student a random number for saving data. We explicitly mentioned that the objective was to evaluate hints, not their programming experience. We stimulated them to ask for as many hints as they wanted, and to request hints when they were starting the next step. This does not necessarily correspond to a real-life situation, but we wanted to gather as many hints as possible. We assume this did not make a difference for the student rating of hints. 


The students worked for 45 minutes on the exercise with the StAP-tutor. We asked them to rate each requested hint.
Our approach was similar to the work of MacNeil \etal~\cite{macneil2023experiences}, where students immediately evaluate explanations after they are shown. After requesting a hint, the interface showed a pop-up with a rating request (see Figure \ref{fig:staptutorrate}). We let students rate hints based on three statements: "The hint is clear", "The hint fits my work", and "The hint is helpful". We chose short questions with a 5-point Likert scale to limit the resistance to rate feedback and limit the time spent on answering. 
We also included the option to add a comment.

After working on the exercise, we asked the students several questions about how it went, how they would compare the tutor to ChatGPT, thoughts on improvements, and overall experience.

\subsubsection{Expert assessment}

We conducted a qualitative evaluation of the hints by two experts: two authors of this paper with extensive experience as a TA and a lecturer, respectively. 
First, we composed a list of nine evaluation criteria based on the literature, see Table~\ref{table: criteria}. For the \textit{Feedback type}, we used Keuning \etal's classification. 
Regarding phrasing, we evaluate the \textit{Tone} and measure the \textit{Length}. We do not consider longer hints to be better since we want to 
keep feedback informative but concise. The criteria \textit{Personalised}, \textit{Appropriate}, and \textit{Misleading Information} directly correlate with the hint's effectiveness. 
Students prefer feedback linked to their work, which we capture in the criterion \textit{Personalised} by checking references to the student's code or approach. We mark hints as \textit{Appropriate} when it fits the current program state. 
\textit{Misleading information} are incorrect statements, which can result in misconceptions. 

\begin{table}[bt]
\caption{Evaluation Criteria.}

\begin{tabularx}{\columnwidth}{p{2cm}X}
\hline
\textbf{Criteria}               & \textbf{Definition}                                                           \\ \hline
\textit{Feedback type}          & What type of feedback is the generated hint?                                             \\

\textit{Information}            & Does the hint contain additional information, such as an explanation, tip or compliment? \\ 

\textit{Level-of-detail}        & Is the hint a bottom-out hint or a high-level description?                                 \\ 

\textit{Personalised}           & Does the hint refer to the student's code or approach?                                   \\ 
\textit{Appropriate}           & Is the hint a suitable next step, given the current state of the student program?        \\ 
\textit{Specific}               & Is the hint limited to only one next step?                                               \\ 
\textit{Misleading information} & Does the hint contain misleading information?                                            \\ 
\textit{Tone}                   & Is the hint phrasing direct, neutral or friendly?                                 \\ 
\textit{Length}                 & What is the length of the hint in sentences?      \\ \hline                                            
\end{tabularx}
\label{table: criteria}
\end{table}


\begin{table*}[tb]
\caption{Inter-rater reliability for the expert evaluation.}
\begin{tabularx}{\linewidth}{lXXXX XXXX}
\hline
    & \textbf{Feedback Type} & \textbf{Information} & \textbf{Level-of-detail} & \textbf{Personalized} & \textbf{Appropriate}  & \textbf{Specific}  & \textbf{Misleading Info }  & \textbf{Tone} \\ \hline
    
Agreements &  19/19   & 18/19   & 16/19   &  18/19                                 & 18/19   & 18/19   & 17/19 & 15/19 \\
Cohen’s kappa   & 0.872 & 0.787 & 0.387 & 0.883    
                    & 0.855  & 0.642  & 0.683 & 0.467  \\
\hline 
\end{tabularx}
\label{table: reliability}
\end{table*}

To verify our interpretations of the categories, we randomly selected 19 hints to compare. Table \ref{table: reliability} shows the agreements and the corresponding inter-rater reliability with Cohen's Kappa. According to Landis \etal's interpretation \cite{landis1977measurement},
\textit{Level-of-detail} had a `fair' and \textit{Tone} a `moderate' score. Because these are more subjective categories, this was not unexpected. Overall, there is substantial agreement, since we disagreed on only 13 of 152 values. We discussed differences, reached a consensus, and one author rated the remaining 29 hints.

%% file: Chapters/Chapter5/Promptengineering.tex
\section{Prompt engineering}\label{chap:prompt}

We performed prompt engineering in an iterative process, where we determined how to proceed based on intermediate findings. In this section, we discuss the results of our design process.

\subsection{Prompt phrasing and attributes}

We started with an exploratory phase in which we tried many different prompts for different programs. From our preprocessed dataset, we selected snapshots from three students working on both assignments. These students used different approaches to solve the exercise, and had different programming competencies.
We selected between 5 and 13 representative snapshots per student sequence.

\begin{figure}[h!]
    \centering
    \includegraphics[width=0.4\textwidth]{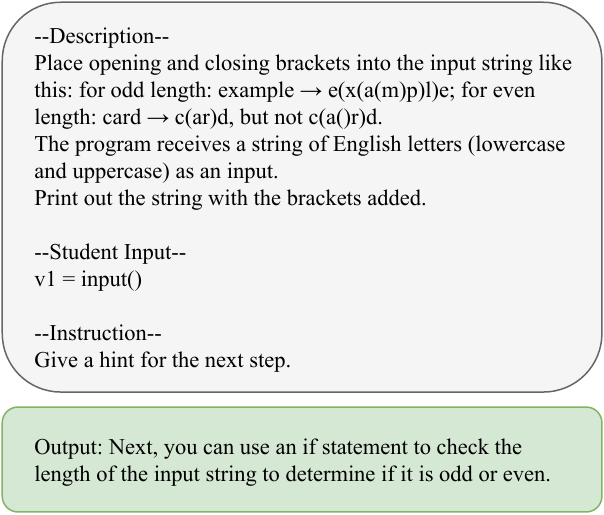}
    \vspace{-5pt}
    \caption{Example prompt with output.}
    \label{fig:exampleprompt}
\end{figure}

The prompt consisted amongst others of the source code for which we wanted to generate a next-step hint, and an instruction describing the goal of the prompt. We first experimented with adding the problem description and model solution attributes to the prompt. Inspired by Sarsa \etal \cite{sarsa2022automatic}, we used formatting to denote the different components of the prompt. Fig.~\ref{fig:exampleprompt} shows an example. 

\noindent We phrased the prompt instruction in three different ways. We included the word \say{step} to specify the essence of what should be generated from the prompt. We created minor variations to examine how adding different keywords such as \say{student} or \say{hint} would affect the generated hints:

\begin{enumerate}
    \item[i)]  “What is the next step?”
    \item[ii)] “Give a hint for the next step.”
    \item[iii)] “Explain the next step for a student”. 
\end{enumerate}

\noindent We combined these 3 instructions with 4 possible combinations of the problem description and model solution attributes: a prompt with no attributes, one of the two and both. This resulted in $3 \cdot 4 = 12$ prompts. For each prompt we generated approximately 60 hints; 10 for each of the 3 students for 2 assignments. We set the temperature to 0.1 to limit the `creativity' of the model and produce somewhat consistent results. The first author of this paper examined all results.

\subsubsection{Problem description and model solution}\label{prompt: att}

We found that only including the problem description resulted in the best outputs. A prompt without the problem description and model solution led to the model guessing what the student could do next, for example:
\begin{quote}
    \say{\textit{The next step likely involves performing some mathematical operation using the values assigned to v2, v3, and v4.}}
\end{quote}
At the start of the exercise, these hints could serve as inspiration on what to do next. However, as the student continued, the model was unable to generate feedback with useful suggestions. 

Adding one of the two attributes caused the next-step hints to be more related to the student code and assignment. However, prompts with a model solution often produced hints that advised students to compare their code to the model solution. For instance:
\begin{quote}
    \say{\textit{Think about how the student's code is different from the model solution. What changes need to be made to the student's code to make it work correctly?}}
\end{quote} 
Despite explicitly asking the model not to refer to the model solution, the model kept doing this. Also, the model would always suggest a step towards the provided model solution, even if a student was solving the problem in another (allowed and correct) way. For instance, an obvious solution for solving the brackets exercise is using a loop. One student who used join operations instead received several `tips' to implement a loop. Or, for the pies exercises, we found students who did not compute the total costs in cents, in contrast to the model solution. As a result, the students received: 
\begin{quote}
    \say{\textit{The student's code seems to be attempting to calculate the cost of buying pies, but it is not following the same approach as the model solution. Try suggesting that the student should use the same formula as the model solution to calculate the total cost of buying pies.}}
\end{quote} 
Already using different variable names would lead to suggestions about changing those names. We tried to rephrase \say{Model solution} as \say{Example solution}, however, this did not solve the problem.
We decided to use prompts with only a problem description. 

\subsubsection{Instructions}

\begin{table*}[bt]
    \centering
\begin{minipage}{0.2\textwidth}
    \includegraphics[width=0.9\linewidth]{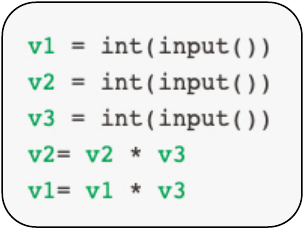}
    \label{fig:piecode}
\end{minipage}%
\begin{minipage}{0.8\textwidth}

\caption{Example hints for the code on the left using different prompt instructions.}

\begin{tabularx}{14cm}{p{2cm}XX}
\hline
\textbf{Prompt} & \textbf{Hint} \\ \hline
\textit{What is the next step}           & The next step is to calculate the total cost of N pies by adding v1 and v2.  \newline    \\ 

\textit{Give a hint for the next step}       & The next step should involve calculating the total cost of N pies using the updated values of v1 and v2, and then converting the total cost into dollars and cents.\newline \vspace{0.0cm}          \\

\textit{Explain the next step for a student} & It seems like you are trying to calculate the total cost of N pies. You have correctly multiplied the cents by the number of pies, but you also need to multiply the dollars by the number of pies. You can do this by updating the value of v1 to be v1 multiplied by v3. Then you can use the updated values of v1 and v2 to calculate the total cost of N pies. \\ \hline 
\end{tabularx}
\label{table: instructions}

\end{minipage}

\end{table*}

We experimented with including different instructions in the prompt. We started with three different instruction phrasings (i--iii), shown in Table \ref{table: instructions} with corresponding hints for an example student program. 
Although the instructions had only slight variations, the generated feedback had notably different results, especially regarding additional information and phrasing. The result of the prompt \say{What is the next step?} most frequently contained code, 
and the phrasing was a bit more straightforward and blunt.
Consequently, we considered this prompt less suitable.



The prompt “Explain the next step for a student” resulted in relatively longer hints, while sometimes also explaining the student's code. Although explanations can help students, we only want to generate hints explaining how to proceed. The phrasing of the generated feedback with this prompt was similar to “Give a hint for the next step”. Both yielded more friendly and carefully phrased hints. Furthermore, this instruction produced feedback formulated as a hint without revealing the exact answer, as in \say{The next step involves computing the total costs in cents.} 

Based on these observations, we combine the keywords from the last two prompts, \say{hint} and \say{student}, to design new instructions.
We omit \say{explain} as it caused long feedback and undesirable code explanations. When trying the prompt: \say{Give this student a hint for the next step}, we found it often generated long outputs. So, we replaced it with two variations that explicitly instruct the model to create shorter hints. Our new set of instructions were:

\begin{enumerate}
    \item[(ii)] \say{Give a hint for the next step.}
    \item[(iv)] \say{Give this student a short hint for the next step.}
    \item[(v)] \say{Give this student a hint for the next step. The hint should be one or two sentences.}
\end{enumerate}

\subsubsection{Temperature}

The next step focused on analysing the impact of the \textit{temperature} parameter. We investigated values from 0.1 to 0.9, with intermediate increments of 0.2 for 
instruction (ii). We briefly compared prompts with only a description and a description combined with a model solution. We wanted to rule out that increasing the temperature would overcome the issues mentioned in section \ref{prompt: att}. Indeed, 
we still encountered the same problems and thus chose to continue only with the description prompt. 

As the value for the temperature increased, we noted that hints contained more unusual or unexpected suggestions. For example, we saw recommendations for using a stack in the brackets exercise, which is unnecessary and probably an unknown data structure to novice programmers. However, we saw improvements too, where hints were more tailored to the students code. We compared the number of useful and unhelpful suggestions. We noticed a shift from having more disadvantages than advantages for every prompt around a temperature of 0.7. Hence, we picked a temperature of 0.5 for our final prompts. 

\subsubsection{Choosing the best prompt}
To choose the best prompt, we used the three prompts from the previous section with a temperature value of 0.5. We randomly picked ten student programs for both exercises. Then, two authors compared the hints generated for each program and ranked them from 1 (best) to 3 (worst). The final score for each prompt was the sum of its total ratings, as shown in table \ref{table: rating}. We selected the prompt with the best total score (v). 

\begin{table}[tb]
\caption{Prompt ranking.}
\begin{tabularx}{0.9\columnwidth}{lXXX} \hline 
 & Prompt (ii) & Prompt (iv) & Prompt (v) \\  \hline

\textbf{Brackets} & 20 & 27 & \textbf{16} \\
\textbf{Pies} & 25 & 19 & \textbf{17} \\ \hline
\textbf{Total score      }       & 45& 46                                                                                           & \textbf{33}       \\
\hline
\end{tabularx}
\label{table: rating}
\end{table}

\subsection{Best prompt practices}

We conclude this section by reflecting on \textbf{SQ1}: What prompt characteristics are suitable for generating effective next-step hints with LLMs?
We found that providing more information, such as a model solution, doesn't always improve the output. LLMs give hints that relate strongly to such a solution, while teachers would be able to recognise alternative solutions and help students without forcing them to change their approach. 
A related issue was that LLMs often explicitly refer to the model solution, even when instructed not to do so. Hellas \etal~\cite{hellas2023exploring} noticed this problem as well. When we included the sentence \say{..use at most three sentences} in the prompt, the output would have \textit{exactly} three sentences. We believe that LLMs cannot handle or interpret certain constraints very well. 

The presentation of feedback is important; we found that keywords such as \say{hint} and \say{student} often help with personal references, explanations, a friendly tone, and compliments. For example:
\begin{quote}
    \say{\textit{The formula for calculating the cost of N pies is correct, but you need to separate the result into dollars and cents. Remember that 100 cents make 1 dollar. You can use the modulo operator (\%) to get the remainder when dividing by 100, which gives the cent value.}}.
\end{quote}
Additionally, using the keyword \say{hint} prevented output that gave away the answer, as we saw with prompts \say{What is the next step?}, which often produced code, while \say{Give a hint for the next step} gave tips such as \say{the next step involves using a for a loop.}

The instruction should be carefully formulated. For example, the model did not recognise when a student completed the assignment and would still recommend a `next step'. We briefly experimented with adding something to the prompt to overcome this problem, such as \say{If the student is done, give a compliment. Else, give a next step hint.} The hints for this prompt would recognise situations where the model was incorrect before. However, the feedback generated was much longer and had (too) many compliments. Adding \say{give a compliment} changed every output where we had expected this to happen for only some of the hints. 

Finally, increasing the temperature value might help produce better feedback. However, this also increases the risk of unusual or useless suggestions, which can be harmful, especially for novices. 
More experienced programmers could probably get inspiration from such feedback and recognise unhelpful hints. 
Increasing the temperature also resulted in more references to the student's code and approach. Choosing this value depends on the context and should be determined through experimentation.


%% file: Chapters/Chapter6/Genertingnextstephints.tex
\section{Evaluating next-step hints}
\label{sec:evaluation}

\subsection{StAP-tutor}
After discovering and choosing the best prompts, we incorporated our findings in our StAP-tutor (\textbf{St}ep \textbf{A}ssisted \textbf{P}rogramming Tutor). The StAP-tutor is a web interface combined with a hint-generation module. The interface has various functionalities, such as choosing exercises, requesting hints and checking your solution, as shown in Fig.~\ref{fig:staptutorrate}. After starting an assignment, the student can ask for help by pressing the hint button. This button sends a request to the hint-generating module, an API written in Python. This API constructs a prompt for the current student program and calls the OpenAI API with the \textbf{gpt-3.5-turbo} model. Then, the API sends the hint back to the interface, where it is displayed to the student.


\subsection{Student experiment}
\begin{figure}[tb]
        \centering
        \includegraphics[width=1\columnwidth]{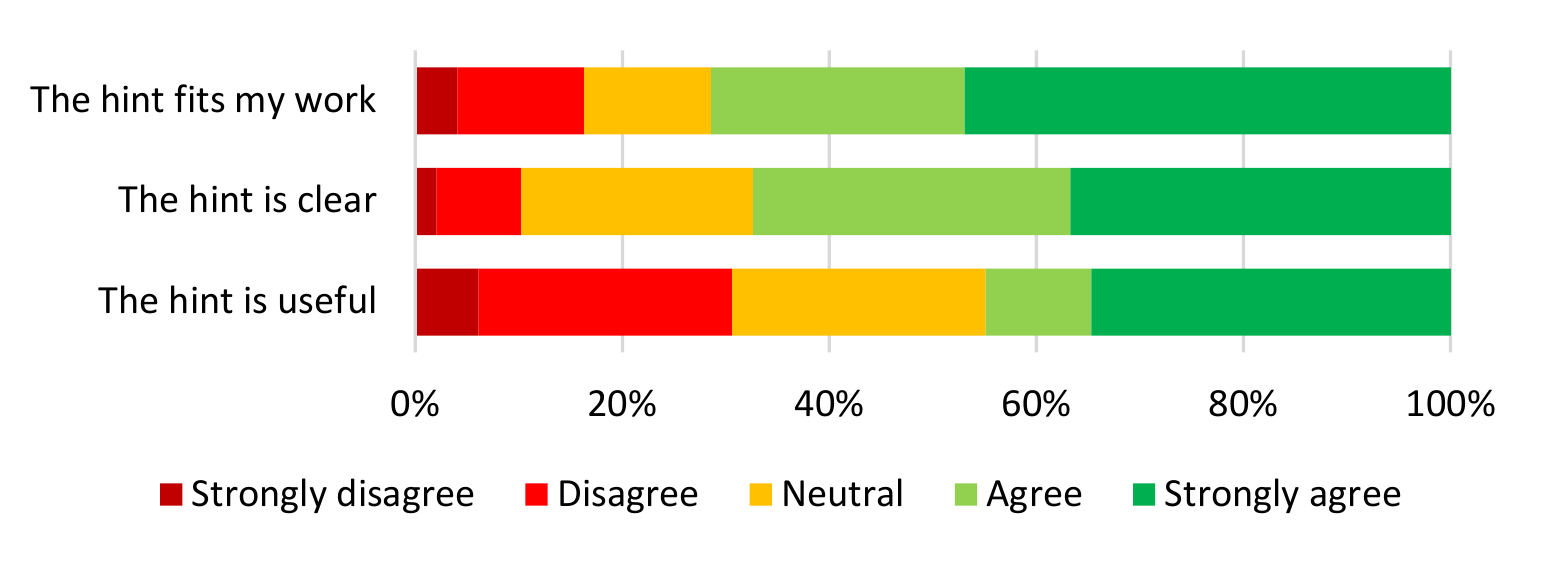}
        \caption{Student hint ratings (n=48).}
        \label{fig: ratings}
\end{figure}

In the experiment, the three students requested 11, 20 and 17 hints, in total 48. The student's hint ratings are in Figure \ref{fig: ratings}. We notice they often marked the hints as clear and suitable for their work. However, they were less convinced about the usefulness of the feedback. Only about half of the time, they tagged the hints as helpful. Some students included a comment with an explanation. One reason was that the feedback was useful, as it was the same as at the beginning of the exercise, but the student had already made significant progress. Another student commented that the hint contradicted a previously given hint, suggesting that the work done did not contribute to the solution.

Even though all students stated that the exercise was aligned with their current skills and programming level, no student finished with a correct solution. They particularly struggled with the logic for correctly counting the clumps, which required tracking and implementing when to `count' a clump with conditional statements. 

After the experiment, we asked the students about their experiences and opinions. Overall, they had a positive impression of the tutor and appreciated the hints, especially at the start of the assignment. At the start, the feedback fitted their work and had practical suggestions for what to do next. However, after a while, when the student was stuck with more complex code, the hints were not helpful anymore. The students needed more detail or practical suggestions, but the feedback was too vague. 

All students mentioned requesting multiple hints for the same code. As the LLM is not deterministic, it might help to regenerate a hint. Students reported that regenerating hints returned a different phrasing or provides more or new information. However, one of the students stated that returning similar hints for the same code could even be more helpful if those hints entail more detailed content. 

Finally, the students agreed that compared to ChatGPT, the StAP-tutor focuses more on your existing code instead of providing a general solution with quite a lot of additional information. The students thought this might help with learning as this also reduces the temptation to copy and paste the assignment and generate a whole solution. However, as one of the students pointed out, an advantage of ChatGPT is that there is a possibility to ask more specific questions yourself. 

\subsection{Expert assessment}

\begin{figure}[tb]
  \centering
    \includegraphics[width=0.8\linewidth]{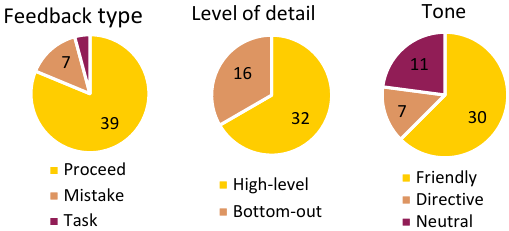}
    \caption{Hint characteristics.}
    \label{fig:fbtyp}
\end{figure}

All 48 generated hints were classified by one or two experts.
Figure \ref{fig:fbtyp} shows the frequencies for the criteria \textit{Feedback Type}, \textit{Level-of-detail} and \textit{Tone}. The majority of the generated hints correspond with the type \textit{Knowledge About How To Proceed}. We found that the hints are more often high-level than bottom-out. Note, in contrast with the type of feedback, we did not state the desired level of detail in the prompt. 
Finally, we found that the feedback mostly had a friendly tone. We interpreted the tone as friendly when the hints were more suggestive (\eg \say{Consider trying to change the condition of the for loop}) than directive (\eg \say{Change the for loop condition}). 

\begin{figure}[tb]
    \centering
    \includegraphics[width=0.45\textwidth]{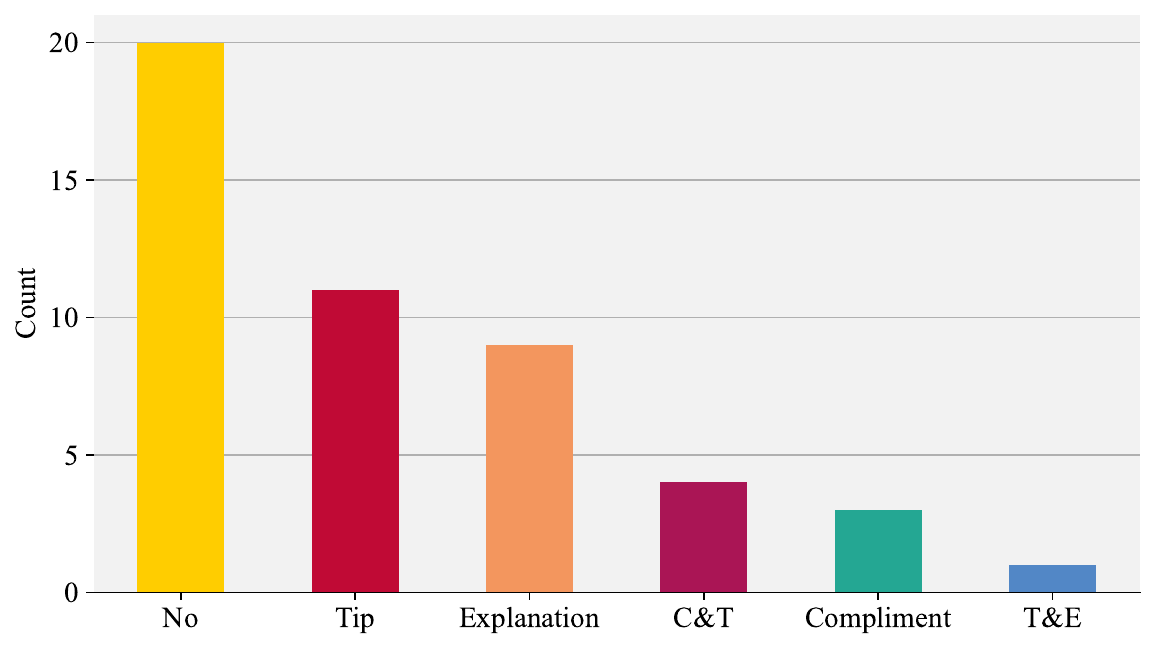}
    \caption{Counts for additional information. C\&T = compliment and tip, T\&E = tip with explanation.}
    \label{fig: information}
\end{figure}

We found that friendly hints often contain additional information. 
Figure \ref{fig: information} shows the frequencies for every additional information category. We occasionally noticed that feedback contained smaller tips in addition to the `main' hint, which often referred to something the student was not yet working on. For example, 
\say{Also, try to think about the conditions under which a clump exists.} or \say{Don't forget to handle the cases where the clump ends and a new clump begins}. 

\begin{figure}[tb]
    \centering
    \includegraphics[width=0.45\textwidth]{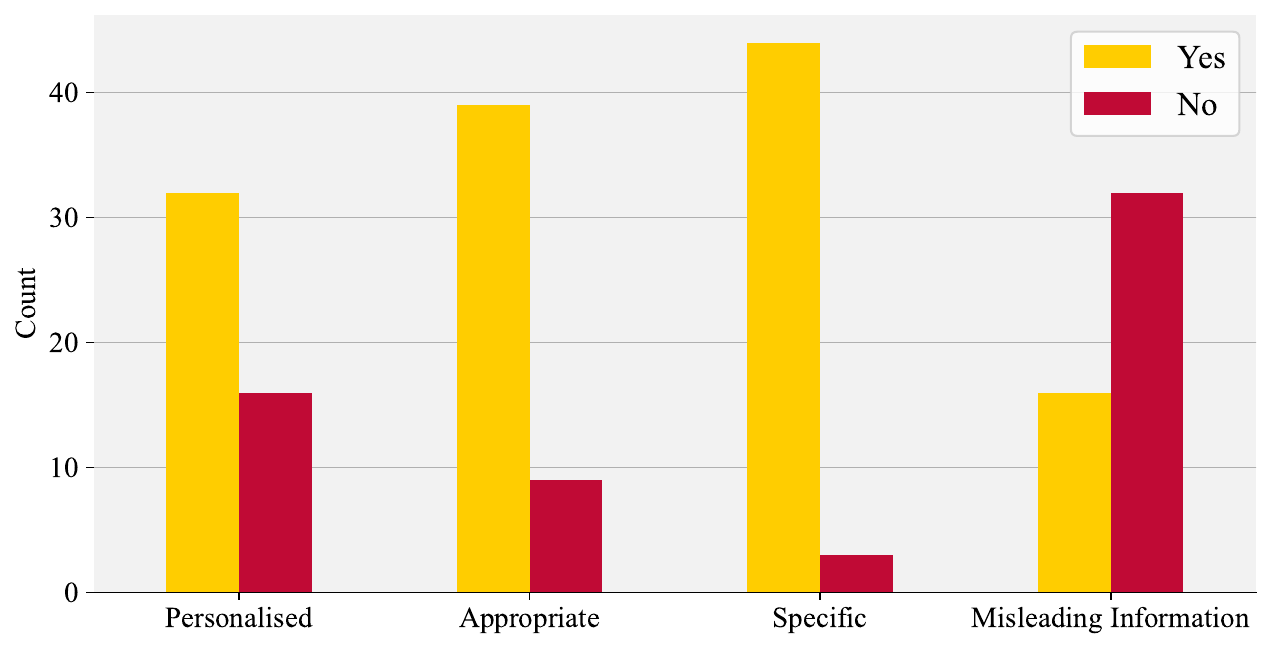}
    \caption{Frequencies of binary evaluation metrics.}
    \label{fig:binaryeval}
\end{figure}

All evaluation metrics with binary values are presented in Figure \ref{fig:binaryeval}. Feedback is frequently personalised, appropriate and specific. We saw hints with explicit references to variables from the student's code and suggestions referring to their current implementation. Furthermore, the feedback usually entailed only one specific step, which was appropriate regarding the student's progress. However, we should be cautious about hints containing misleading information. We noted the misleading information was not notably wrong at first sight, but often more subtle. For example, feedback explained incorrectly why some piece of code is incorrect: \say{The for loop of your current code may reach an out-of-range index error if it reaches the end of the list without finding a clump}. Or, it gives suggestions that will not help to solve the problem: \say{You need to make sure your program doesn't break when the input array has only one element.} Finally, we found misleading information is often connected to hints that refer to edge cases, while a correct solution did not require handling these cases separately.

\subsection{Conclusion}

In this section we answer \textbf{SQ2}: What are students' and experts' perceptions of the quality of LLM-generated next-step hints, and how do they rate them?
We found that both students and experts are positive about the hints. While the feedback provided concise and tailored support, it was not always helpful and sometimes contained misleading information. Solving programming exercises entails breaking down a problem into smaller sub-tasks, and next-step hints may help students performing a sequence of steps leading to a solution.
 

Overall, the students found the hints clear. Most feedback consisted of 1 or 2 sentences, with some outliers to 3. We think LLM-generated feedback from these prompts has the right level of complexity and length. LLMs respond with appropriate and potentially helpful answers, as in the work of Hellas \etal \cite{hellas2023exploring}. Students sometimes ask for hints multiple times for the same code repeatedly.


All students indicated they needed more support than the generated next-step hints. 
The prompt we used generated mainly high-level hints. These hints may help the students get started and support them with constructing an idea for an overall approach. Yet, when approaching the end of the exercise, students want more detailed feedback related to their solution. 

The feedback frequently included additional information, such as compliments, tips and explanations. Students liked the `motivational words', which made the interaction feel more personal. 
We found it important that the feedback was not overly confident and directive, to encourage a student to think critically about the hint.

An obvious limitation is that hints occasionally contain misleading information. This issue is hard to resolve with further prompt engineering as we believe it is more related to the overall capability of LLMs themselves. We expect that as the LLM performance increases, the amount of misleading information will decrease. When using LLMs for teaching, we should be cautious and warn students about them providing incorrect information.

%% file: Chapters/Chapter7/Discussion.tex
\section{Discussion}
\label{sec:discussion}
This section reviews our main research question and discusses the limitations of our study regarding prompt-engineering, the experimental set-up, and reproducibility. We discuss how follow-up research may address these limitations. 

\subsection{Generating next-step hints using LLMs}

Answering \textbf{RQ1}, "To what extent can we use LLMs to generate informative and effective next-step hints for Python introductory exercises?",
we found it is best to use prompts that refer to a problem description and include keywords related to the application context, such as \say{student} or \say{hint}. Compared to other data-driven approaches, LLMs can generate personalised, tailored feedback, while they do not require large datasets or complex computational methods. In contrast with standard hints, which often are mere code edit suggestions, these hints contain compliments, explanations and tips. This additional information was well-received by the students. Nevertheless, there are still some points for improvement, as LLM-generated feedback may entail misleading information. 

\subsection{Threats to validity}
\subsubsection{Prompt engineering}

Although we extensively experimented with engineering prompts, we had to limit ourselves. First, we primarily examined the effects of using different attributes, phrasings and temperature values isolated from each other. This approach allowed us to analyse the influence of these characteristics independently. These factors may influence each other, but investigating all possible combinations would require too much time.
Second, we only analysed a small set of prompt instructions. As we were content with the preliminary results, we chose not to explore different phrasings, which might have led to increased hints quality. 
Lastly, we did not use strict evaluation criteria in the prompt engineering phase.
Because we wanted to get a general impression of how the LLM performed for a wide variety of code states, we did not generate multiple hints for one student program.
\subsubsection{Experimental setup}
We performed the experiments with only three students. Furthermore, we only examined their opinions instead of factors like learning gain. Therefore, we cannot make strong statements about the effect of the hints. Such conclusions require a long-term study with a large group of students and pre- and post-tests with a control group. 

In our expert assessment, we tried to capture most important factors for effective feedback in our evaluation criteria. Nevertheless, capturing effectiveness based on a few characteristics is challenging since this depends on many factors. For instance, students could ask for hints for themselves, which is not necessarily proven to be the best practice \cite{jeuring2022towards}. Implementing other methods for timing the feedback, however, require a method for automating hint delivery, which was outside the scope of this work. 

\subsubsection{Reproducibility} 
A known issue with doing LLM-studies is that models are constantly updated, which might produce other results. During our experiments, OpenAI released GPT-4. We performed some minor experiments using GPT-4 and compared its results with GPT-3.5-turbo. We found that the same prompt for both models outputs, at first glance, different results. 
GPT-4 created hints with good suggestions we had not seen before with our best prompts. We expect our prompt practices also work for updated versions, but other research is required to confirm this.

\subsection{Future work}\label{future}
Although not perfect, at this time, LLM-generated hints certainly have potential. We propose various avenues for future work to improve generating (next-step) hints with LLMs.  

Students may benefit from help at different levels depending on the context. Once students are stuck, they might require more in-depth feedback that offers better guidance than high-level hints. 
Our StAP-tutor uses only one prompt for hint generation, which generates both bottom-out and high-level hints. 
We could further investigate the relation between prompts and the level of detail of the hints they generate. 
These findings could be combined with advanced methods for student modelling to provide personalised feedback and guidance based on a student's progress. For instance, depending on the number of hints requested for a specific state, a different prompt could be used to obtain more guided feedback. 

Students pointed out that they could not influence the topic of hints. 
Prather \etal found similar frustrations by students while working with Copilot \cite{prather2023s}. 
Hints occasionally did not correspond to the part of the code that students were struggling with. For example, the StAP-tutor suggested writing already-written pieces of code. We propose to investigate student control, so students can indicate what part of the code they want help with, and to instruct the model to give more specific feedback.

Other ideas are adding few-shot learning or experimenting with giving the prompt multiple solutions. With few-shot-learning, examples are included in the prompts \cite{brown2020language}, from which the LLMs could learn and follow its structure. However, this requires a set of expert-validated example hints, which we did not have for our dataset. 

%% file: Chapters/Chapter7/Conclusion.tex
\section{Conclusion}
\label{sec:conclusions}
We investigated how to use LLMs to generate next-step hints for introductory programming exercises in Python. These hints should support students in their learning, and not give away the entire solution. First, we explored various prompt practices to discover what prompts would yield the best feedback. We analysed the effects of adding a model solution and description to the prompt, using different instructions and varying the temperature. We found it is best to only include a problem description and use keywords such as \say{student} and \say{hint} in the prompt. In addition, increasing the temperature parameter might contribute to more personalised feedback. However, too high values may cause unhelpful suggestions.

We then created the StAP-tutor based on these results. With this tool, students can ask for hints while doing programming exercises. Students working with the tutor had an overall positive impression of the hints. The LLM-generated feedback was personalised, appropriate and contained helpful additional information such as tips and explanations. Our expert assessment support these findings, but also found that the hints contained misleading information. Another issue was that the high-level feedback did not support students at every stage of their problem-solving.
Our study shows the potential of using LLMs in programming education by generating hints, and proposes several areas for future work.

%% file: main.bbl

\begin{thebibliography}{42}


\ifx \showCODEN    \undefined \def \showCODEN     #1{\unskip}     \fi
\ifx \showDOI      \undefined \def \showDOI       #1{#1}\fi
\ifx \showISBNx    \undefined \def \showISBNx     #1{\unskip}     \fi
\ifx \showISBNxiii \undefined \def \showISBNxiii  #1{\unskip}     \fi
\ifx \showISSN     \undefined \def \showISSN      #1{\unskip}     \fi
\ifx \showLCCN     \undefined \def \showLCCN      #1{\unskip}     \fi
\ifx \shownote     \undefined \def \shownote      #1{#1}          \fi
\ifx \showarticletitle \undefined \def \showarticletitle #1{#1}   \fi
\ifx \showURL      \undefined \def \showURL       {\relax}        \fi
\providecommand\bibfield[2]{#2}
\providecommand\bibinfo[2]{#2}
\providecommand\natexlab[1]{#1}
\providecommand\showeprint[2][]{arXiv:#2}

\bibitem[Anderson and Reiser(1985)]%
        {anderson1985lisp}
\bibfield{author}{\bibinfo{person}{John~R Anderson} {and}
  \bibinfo{person}{Brian~J Reiser}.} \bibinfo{year}{1985}\natexlab{}.
\newblock \showarticletitle{The LISP tutor}.
\newblock \bibinfo{journal}{\emph{Byte}} \bibinfo{volume}{10},
  \bibinfo{number}{4} (\bibinfo{year}{1985}), \bibinfo{pages}{159--175}.
\newblock


\bibitem[Becker et~al\mbox{.}(2023)]%
        {becker2023programming}
\bibfield{author}{\bibinfo{person}{Brett~A Becker}, \bibinfo{person}{Paul
  Denny}, \bibinfo{person}{James Finnie-Ansley}, \bibinfo{person}{Andrew
  Luxton-Reilly}, \bibinfo{person}{James Prather}, {and}
  \bibinfo{person}{Eddie~Antonio Santos}.} \bibinfo{year}{2023}\natexlab{}.
\newblock \showarticletitle{Programming Is Hard-Or at Least It Used to Be:
  Educational Opportunities and Challenges of AI Code Generation}. In
  \bibinfo{booktitle}{\emph{Proc. of SIGCSE}}. \bibinfo{pages}{500--506}.
\newblock


\bibitem[Brown et~al\mbox{.}(2020)]%
        {brown2020language}
\bibfield{author}{\bibinfo{person}{Tom Brown}, \bibinfo{person}{Benjamin Mann},
  \bibinfo{person}{Nick Ryder}, \bibinfo{person}{Melanie Subbiah},
  \bibinfo{person}{Jared~D Kaplan}, \bibinfo{person}{Prafulla Dhariwal},
  \bibinfo{person}{Arvind Neelakantan}, \bibinfo{person}{Pranav Shyam},
  \bibinfo{person}{Girish Sastry}, \bibinfo{person}{Amanda Askell},
  {et~al\mbox{.}}} \bibinfo{year}{2020}\natexlab{}.
\newblock \showarticletitle{Language models are few-shot learners}.
\newblock \bibinfo{journal}{\emph{Advances in neural information processing
  systems}}  \bibinfo{volume}{33} (\bibinfo{year}{2020}),
  \bibinfo{pages}{1877--1901}.
\newblock


\bibitem[Chen et~al\mbox{.}(2021)]%
        {chen2021evaluating}
\bibfield{author}{\bibinfo{person}{Mark Chen}, \bibinfo{person}{Jerry Tworek},
  \bibinfo{person}{Heewoo Jun}, \bibinfo{person}{Qiming Yuan},
  \bibinfo{person}{Henrique Ponde de~Oliveira Pinto}, \bibinfo{person}{Jared
  Kaplan}, \bibinfo{person}{Harri Edwards}, \bibinfo{person}{Yuri Burda},
  \bibinfo{person}{Nicholas Joseph}, \bibinfo{person}{Greg Brockman},
  {et~al\mbox{.}}} \bibinfo{year}{2021}\natexlab{}.
\newblock \showarticletitle{Evaluating large language models trained on code}.
\newblock \bibinfo{journal}{\emph{arXiv preprint arXiv:2107.03374}}
  (\bibinfo{year}{2021}).
\newblock


\bibitem[Dawson et~al\mbox{.}(2019)]%
        {dawson2019makes}
\bibfield{author}{\bibinfo{person}{Phillip Dawson}, \bibinfo{person}{Michael
  Henderson}, \bibinfo{person}{Paige Mahoney}, \bibinfo{person}{Michael
  Phillips}, \bibinfo{person}{Tracii Ryan}, \bibinfo{person}{David Boud}, {and}
  \bibinfo{person}{Elizabeth Molloy}.} \bibinfo{year}{2019}\natexlab{}.
\newblock \showarticletitle{What makes for effective feedback: Staff and
  student perspectives}.
\newblock \bibinfo{journal}{\emph{Assessment \& Evaluation in Higher Ed.}}
  \bibinfo{volume}{44}, \bibinfo{number}{1} (\bibinfo{year}{2019}),
  \bibinfo{pages}{25--36}.
\newblock


\bibitem[Deeva et~al\mbox{.}(2021)]%
        {deeva2021review}
\bibfield{author}{\bibinfo{person}{Galina Deeva}, \bibinfo{person}{Daria
  Bogdanova}, \bibinfo{person}{Estefan{\'\i}a Serral}, \bibinfo{person}{Monique
  Snoeck}, {and} \bibinfo{person}{Jochen De~Weerdt}.}
  \bibinfo{year}{2021}\natexlab{}.
\newblock \showarticletitle{A review of automated feedback systems for
  learners: Classification framework, challenges and opportunities}.
\newblock \bibinfo{journal}{\emph{Computers \& Education}}
  \bibinfo{volume}{162} (\bibinfo{year}{2021}).
\newblock


\bibitem[Denny et~al\mbox{.}(2023a)]%
        {denny2023conversing}
\bibfield{author}{\bibinfo{person}{Paul Denny}, \bibinfo{person}{Viraj Kumar},
  {and} \bibinfo{person}{Nasser Giacaman}.} \bibinfo{year}{2023}\natexlab{a}.
\newblock \showarticletitle{Conversing with copilot: Exploring prompt
  engineering for solving cs1 problems using natural language}. In
  \bibinfo{booktitle}{\emph{Proc. of SIGCSE}}. \bibinfo{pages}{1136--1142}.
\newblock


\bibitem[Denny et~al\mbox{.}(2023b)]%
        {denny2023computing}
\bibfield{author}{\bibinfo{person}{Paul Denny}, \bibinfo{person}{James
  Prather}, \bibinfo{person}{Brett~A Becker}, \bibinfo{person}{James
  Finnie-Ansley}, \bibinfo{person}{Arto Hellas}, \bibinfo{person}{Juho
  Leinonen}, \bibinfo{person}{Andrew Luxton-Reilly}, \bibinfo{person}{Brent~N
  Reeves}, \bibinfo{person}{Eddie~Antonio Santos}, {and} \bibinfo{person}{Sami
  Sarsa}.} \bibinfo{year}{2023}\natexlab{b}.
\newblock \showarticletitle{Computing Education in the Era of Generative AI}.
\newblock \bibinfo{journal}{\emph{arXiv preprint arXiv:2306.02608}}
  (\bibinfo{year}{2023}).
\newblock


\bibitem[D{\"o}derlein et~al\mbox{.}(2022)]%
        {doderlein2022piloting}
\bibfield{author}{\bibinfo{person}{Jean-Baptiste D{\"o}derlein},
  \bibinfo{person}{Mathieu Acher}, \bibinfo{person}{Djamel Eddine~Khelladi},
  {and} \bibinfo{person}{Benoit Combemale}.} \bibinfo{year}{2022}\natexlab{}.
\newblock \showarticletitle{Piloting Copilot and Codex: Hot Temperature, Cold
  Prompts, or Black Magic?}
\newblock \bibinfo{journal}{\emph{arXiv e-prints}} (\bibinfo{year}{2022}).
\newblock


\bibitem[Finnie-Ansley et~al\mbox{.}(2022)]%
        {finnie2022robots}
\bibfield{author}{\bibinfo{person}{James Finnie-Ansley}, \bibinfo{person}{Paul
  Denny}, \bibinfo{person}{Brett~A Becker}, \bibinfo{person}{Andrew
  Luxton-Reilly}, {and} \bibinfo{person}{James Prather}.}
  \bibinfo{year}{2022}\natexlab{}.
\newblock \showarticletitle{The Robots Are Coming: Exploring the Implications
  of OpenAI Codex on Introductory Programming}. In
  \bibinfo{booktitle}{\emph{Proc. of ACE}}. \bibinfo{pages}{10--19}.
\newblock


\bibitem[Finnie-Ansley et~al\mbox{.}(2023)]%
        {finnie2023my}
\bibfield{author}{\bibinfo{person}{James Finnie-Ansley}, \bibinfo{person}{Paul
  Denny}, \bibinfo{person}{Andrew Luxton-Reilly},
  \bibinfo{person}{Eddie~Antonio Santos}, \bibinfo{person}{James Prather},
  {and} \bibinfo{person}{Brett~A Becker}.} \bibinfo{year}{2023}\natexlab{}.
\newblock \showarticletitle{My AI Wants to Know if This Will Be on the Exam:
  Testing OpenAI’s Codex on CS2 Programming Exercises}. In
  \bibinfo{booktitle}{\emph{Proc. of ACE}}. \bibinfo{pages}{97--104}.
\newblock


\bibitem[Hao et~al\mbox{.}(2022)]%
        {hao2022towards}
\bibfield{author}{\bibinfo{person}{Qiang Hao}, \bibinfo{person}{David~H
  Smith~IV}, \bibinfo{person}{Lu Ding}, \bibinfo{person}{Amy Ko},
  \bibinfo{person}{Camille Ottaway}, \bibinfo{person}{Jack Wilson},
  \bibinfo{person}{Kai~H Arakawa}, \bibinfo{person}{Alistair Turcan},
  \bibinfo{person}{Timothy Poehlman}, {and} \bibinfo{person}{Tyler Greer}.}
  \bibinfo{year}{2022}\natexlab{}.
\newblock \showarticletitle{Towards understanding the effective design of
  automated formative feedback for programming assignments}.
\newblock \bibinfo{journal}{\emph{Computer Science Education}}
  \bibinfo{volume}{32}, \bibinfo{number}{1} (\bibinfo{year}{2022}),
  \bibinfo{pages}{105--127}.
\newblock


\bibitem[Hattie and Timperley(2007)]%
        {hattie2007power}
\bibfield{author}{\bibinfo{person}{John Hattie} {and} \bibinfo{person}{Helen
  Timperley}.} \bibinfo{year}{2007}\natexlab{}.
\newblock \showarticletitle{The power of feedback}.
\newblock \bibinfo{journal}{\emph{Review of educational research}}
  \bibinfo{volume}{77}, \bibinfo{number}{1} (\bibinfo{year}{2007}),
  \bibinfo{pages}{81--112}.
\newblock


\bibitem[Hellas et~al\mbox{.}(2023)]%
        {hellas2023exploring}
\bibfield{author}{\bibinfo{person}{Arto Hellas}, \bibinfo{person}{Juho
  Leinonen}, \bibinfo{person}{Sami Sarsa}, \bibinfo{person}{Charles Koutcheme},
  \bibinfo{person}{Lilja Kujanp\"{a}\"{a}}, {and} \bibinfo{person}{Juha
  Sorva}.} \bibinfo{year}{2023}\natexlab{}.
\newblock \showarticletitle{Exploring the Responses of Large Language Models to
  Beginner Programmers’ Help Requests}. In \bibinfo{booktitle}{\emph{Proc. of
  ICER}}. \bibinfo{pages}{93–105}.
\newblock


\bibitem[Irons and Elkington(2021)]%
        {irons2021enhancing}
\bibfield{author}{\bibinfo{person}{Alastair Irons} {and} \bibinfo{person}{Sam
  Elkington}.} \bibinfo{year}{2021}\natexlab{}.
\newblock \bibinfo{booktitle}{\emph{Enhancing learning through formative
  assessment and feedback}}.
\newblock \bibinfo{publisher}{Routledge}.
\newblock


\bibitem[Jeuring et~al\mbox{.}(2022)]%
        {jeuring2022towards}
\bibfield{author}{\bibinfo{person}{Johan Jeuring}, \bibinfo{person}{Hieke
  Keuning}, \bibinfo{person}{Samiha Marwan}, \bibinfo{person}{Dennis Bouvier},
  \bibinfo{person}{Cruz Izu}, \bibinfo{person}{Natalie Kiesler},
  \bibinfo{person}{Teemu Lehtinen}, \bibinfo{person}{Dominic Lohr},
  \bibinfo{person}{Andrew Peterson}, {and} \bibinfo{person}{Sami Sarsa}.}
  \bibinfo{year}{2022}\natexlab{}.
\newblock \showarticletitle{Towards Giving Timely Formative Feedback and Hints
  to Novice Programmers}.
\newblock In \bibinfo{booktitle}{\emph{ITiCSE Working Group Reports}}.
  \bibinfo{pages}{95--115}.
\newblock


\bibitem[Kazemitabaar et~al\mbox{.}(2023)]%
        {kazemitabaar2023studying}
\bibfield{author}{\bibinfo{person}{Majeed Kazemitabaar},
  \bibinfo{person}{Justin Chow}, \bibinfo{person}{Carl Ka~To Ma},
  \bibinfo{person}{Barbara~J Ericson}, \bibinfo{person}{David Weintrop}, {and}
  \bibinfo{person}{Tovi Grossman}.} \bibinfo{year}{2023}\natexlab{}.
\newblock \showarticletitle{Studying the effect of AI Code Generators on
  Supporting Novice Learners in Introductory Programming}. In
  \bibinfo{booktitle}{\emph{Proc. of the CHI Conference}}.
  \bibinfo{pages}{1--23}.
\newblock


\bibitem[Keuning et~al\mbox{.}(2014)]%
        {keuning2014strategy}
\bibfield{author}{\bibinfo{person}{Hieke Keuning}, \bibinfo{person}{Bastiaan
  Heeren}, {and} \bibinfo{person}{Johan Jeuring}.}
  \bibinfo{year}{2014}\natexlab{}.
\newblock \showarticletitle{Strategy-based feedback in a programming tutor}. In
  \bibinfo{booktitle}{\emph{Proc. of the Computer Science Education Research
  Conference}}. \bibinfo{pages}{43--54}.
\newblock


\bibitem[Keuning et~al\mbox{.}(2018)]%
        {keuning2018systematic}
\bibfield{author}{\bibinfo{person}{Hieke Keuning}, \bibinfo{person}{Johan
  Jeuring}, {and} \bibinfo{person}{Bastiaan Heeren}.}
  \bibinfo{year}{2018}\natexlab{}.
\newblock \showarticletitle{A systematic literature review of automated
  feedback generation for programming exercises}.
\newblock \bibinfo{journal}{\emph{ACM TOCE}} \bibinfo{volume}{19},
  \bibinfo{number}{1} (\bibinfo{year}{2018}), \bibinfo{pages}{1--43}.
\newblock


\bibitem[Kiesler et~al\mbox{.}(2023)]%
        {kiesler2023exploring}
\bibfield{author}{\bibinfo{person}{Natalie Kiesler}, \bibinfo{person}{Dominic
  Lohr}, {and} \bibinfo{person}{Hieke Keuning}.}
  \bibinfo{year}{2023}\natexlab{}.
\newblock \showarticletitle{Exploring the Potential of Large Language Models to
  Generate Formative Programming Feedback}.
\newblock \bibinfo{journal}{\emph{arXiv preprint arXiv:2309.00029}}
  (\bibinfo{year}{2023}).
\newblock


\bibitem[Landis and Koch(1977)]%
        {landis1977measurement}
\bibfield{author}{\bibinfo{person}{J~Richard Landis} {and}
  \bibinfo{person}{Gary~G Koch}.} \bibinfo{year}{1977}\natexlab{}.
\newblock \showarticletitle{The measurement of observer agreement for
  categorical data}.
\newblock \bibinfo{journal}{\emph{Biometrics}} (\bibinfo{year}{1977}),
  \bibinfo{pages}{159--174}.
\newblock


\bibitem[Leinonen et~al\mbox{.}(2023)]%
        {leinonen2023using}
\bibfield{author}{\bibinfo{person}{Juho Leinonen}, \bibinfo{person}{Arto
  Hellas}, \bibinfo{person}{Sami Sarsa}, \bibinfo{person}{Brent Reeves},
  \bibinfo{person}{Paul Denny}, \bibinfo{person}{James Prather}, {and}
  \bibinfo{person}{Brett~A Becker}.} \bibinfo{year}{2023}\natexlab{}.
\newblock \showarticletitle{Using large language models to enhance programming
  error messages}. In \bibinfo{booktitle}{\emph{Proc. of SIGCSE}}.
  \bibinfo{pages}{563--569}.
\newblock


\bibitem[Lyulina et~al\mbox{.}(2021)]%
        {Lyulina2021tasktracker}
\bibfield{author}{\bibinfo{person}{Elena Lyulina}, \bibinfo{person}{Anastasiia
  Birillo}, \bibinfo{person}{Vladimir Kovalenko}, {and}
  \bibinfo{person}{Timofey Bryksin}.} \bibinfo{year}{2021}\natexlab{}.
\newblock \showarticletitle{TaskTracker-Tool: A Toolkit for Tracking of Code
  Snapshots and Activity Data During Solution of Programming Tasks}. In
  \bibinfo{booktitle}{\emph{Proc. of SIGCSE}}. \bibinfo{pages}{495–501}.
\newblock


\bibitem[MacNeil et~al\mbox{.}(2023)]%
        {macneil2023experiences}
\bibfield{author}{\bibinfo{person}{Stephen MacNeil}, \bibinfo{person}{Andrew
  Tran}, \bibinfo{person}{Arto Hellas}, \bibinfo{person}{Joanne Kim},
  \bibinfo{person}{Sami Sarsa}, \bibinfo{person}{Paul Denny},
  \bibinfo{person}{Seth Bernstein}, {and} \bibinfo{person}{Juho Leinonen}.}
  \bibinfo{year}{2023}\natexlab{}.
\newblock \showarticletitle{Experiences from using code explanations generated
  by large language models in a web software development e-book}. In
  \bibinfo{booktitle}{\emph{Proc. of SIGCSE}}. \bibinfo{pages}{931--937}.
\newblock


\bibitem[MacNeil et~al\mbox{.}(2022)]%
        {macneil2022generating}
\bibfield{author}{\bibinfo{person}{Stephen MacNeil}, \bibinfo{person}{Andrew
  Tran}, \bibinfo{person}{Dan Mogil}, \bibinfo{person}{Seth Bernstein},
  \bibinfo{person}{Erin Ross}, {and} \bibinfo{person}{Ziheng Huang}.}
  \bibinfo{year}{2022}\natexlab{}.
\newblock \showarticletitle{Generating diverse code explanations using the
  gpt-3 large language model}. In \bibinfo{booktitle}{\emph{Proc. of ICER}}.
  \bibinfo{pages}{37--39}.
\newblock


\bibitem[Malysheva and Kelleher(2022)]%
        {malysheva2022algorithm}
\bibfield{author}{\bibinfo{person}{Yana Malysheva} {and}
  \bibinfo{person}{Caitlin Kelleher}.} \bibinfo{year}{2022}\natexlab{}.
\newblock \showarticletitle{An Algorithm for Generating Explainable Corrections
  to Student Code}. In \bibinfo{booktitle}{\emph{Proc. of Koli Calling}}.
  \bibinfo{pages}{1--11}.
\newblock


\bibitem[Marwan et~al\mbox{.}(2019)]%
        {marwan2019impact}
\bibfield{author}{\bibinfo{person}{Samiha Marwan}, \bibinfo{person}{Nicholas
  Lytle}, \bibinfo{person}{Joseph~Jay Williams}, {and} \bibinfo{person}{Thomas
  Price}.} \bibinfo{year}{2019}\natexlab{}.
\newblock \showarticletitle{The impact of adding textual explanations to
  next-step hints in a novice programming environment}. In
  \bibinfo{booktitle}{\emph{Proc. of ITiCSE}}. \bibinfo{pages}{520--526}.
\newblock


\bibitem[Mousavinasab et~al\mbox{.}(2021)]%
        {mousavinasab2021intelligent}
\bibfield{author}{\bibinfo{person}{Elham Mousavinasab}, \bibinfo{person}{Nahid
  Zarifsanaiey}, \bibinfo{person}{Sharareh R.~Niakan~Kalhori},
  \bibinfo{person}{Mahnaz Rakhshan}, \bibinfo{person}{Leila Keikha}, {and}
  \bibinfo{person}{Marjan Ghazi~Saeedi}.} \bibinfo{year}{2021}\natexlab{}.
\newblock \showarticletitle{Intelligent tutoring systems: a systematic review
  of characteristics, applications, and evaluation methods}.
\newblock \bibinfo{journal}{\emph{Interactive Learning Environments}}
  \bibinfo{volume}{29}, \bibinfo{number}{1} (\bibinfo{year}{2021}),
  \bibinfo{pages}{142--163}.
\newblock


\bibitem[Narciss(2008)]%
        {narciss2008feedback}
\bibfield{author}{\bibinfo{person}{Susanne Narciss}.}
  \bibinfo{year}{2008}\natexlab{}.
\newblock \showarticletitle{Feedback strategies for interactive learning
  tasks}.
\newblock In \bibinfo{booktitle}{\emph{Handbook of research on educ.
  communications and technology}}. \bibinfo{pages}{125--143}.
\newblock


\bibitem[OpenAI(2023)]%
        {OpenAI2023GPT4TR}
\bibfield{author}{\bibinfo{person}{OpenAI}.} \bibinfo{year}{2023}\natexlab{}.
\newblock \showarticletitle{GPT-4 Technical Report}.
\newblock \bibinfo{journal}{\emph{ArXiv}}  \bibinfo{volume}{abs/2303.08774}
  (\bibinfo{year}{2023}).
\newblock


\bibitem[Paassen et~al\mbox{.}(2018)]%
        {paassen2018continuous}
\bibfield{author}{\bibinfo{person}{Benjamin Paassen}, \bibinfo{person}{Barbara
  Hammer}, \bibinfo{person}{Thomas~W Price}, \bibinfo{person}{Tiffany Barnes},
  \bibinfo{person}{Sebastian Gross}, \bibinfo{person}{Niels Pinkwart},
  {et~al\mbox{.}}} \bibinfo{year}{2018}\natexlab{}.
\newblock \showarticletitle{The Continuous Hint Factory-Providing Hints in Vast
  and Sparsely Populated Edit Distance Spaces}.
\newblock \bibinfo{journal}{\emph{Journal of Educational Data Mining}}
  \bibinfo{volume}{10}, \bibinfo{number}{1} (\bibinfo{year}{2018}),
  \bibinfo{pages}{1--35}.
\newblock


\bibitem[Phung et~al\mbox{.}(2023)]%
        {phung2023generating}
\bibfield{author}{\bibinfo{person}{Tung Phung}, \bibinfo{person}{Jos{\'e}
  Cambronero}, \bibinfo{person}{Sumit Gulwani}, \bibinfo{person}{Tobias Kohn},
  \bibinfo{person}{Rupak Majumdar}, \bibinfo{person}{Adish Singla}, {and}
  \bibinfo{person}{Gustavo Soares}.} \bibinfo{year}{2023}\natexlab{}.
\newblock \showarticletitle{Generating High-Precision Feedback for Programming
  Syntax Errors using Large Language Models}.
\newblock \bibinfo{journal}{\emph{arXiv preprint arXiv:2302.04662}}
  (\bibinfo{year}{2023}).
\newblock


\bibitem[Prather et~al\mbox{.}(2023a)]%
        {prather2023robots}
\bibfield{author}{\bibinfo{person}{James Prather}, \bibinfo{person}{Paul
  Denny}, \bibinfo{person}{Juho Leinonen}, \bibinfo{person}{Brett~A. Becker},
  \bibinfo{person}{Ibrahim Albluwi}, \bibinfo{person}{Michelle Craig},
  \bibinfo{person}{Hieke Keuning}, \bibinfo{person}{Natalie Kiesler},
  \bibinfo{person}{Tobias Kohn}, \bibinfo{person}{Andrew Luxton-Reilly},
  \bibinfo{person}{Stephen MacNeil}, \bibinfo{person}{Andrew Peterson},
  \bibinfo{person}{Raymond Pettit}, \bibinfo{person}{Brent~N. Reeves}, {and}
  \bibinfo{person}{Jaromir Savelka}.} \bibinfo{year}{2023}\natexlab{a}.
\newblock \showarticletitle{The Robots are Here: Navigating the Generative AI
  Revolution in Computing Education}.
\newblock \bibinfo{journal}{\emph{arXiv preprint arXiv:2310.00658}}
  (\bibinfo{year}{2023}).
\newblock


\bibitem[Prather et~al\mbox{.}(2023b)]%
        {prather2023s}
\bibfield{author}{\bibinfo{person}{James Prather}, \bibinfo{person}{Brent~N.
  Reeves}, \bibinfo{person}{Paul Denny}, \bibinfo{person}{Brett~A. Becker},
  \bibinfo{person}{Juho Leinonen}, \bibinfo{person}{Andrew Luxton-Reilly},
  \bibinfo{person}{Garrett Powell}, \bibinfo{person}{James Finnie-Ansley},
  {and} \bibinfo{person}{Eddie~Antonio Santos}.}
  \bibinfo{year}{2023}\natexlab{b}.
\newblock \showarticletitle{``It’s Weird That It Knows What I Want'':
  Usability and Interactions with Copilot for Novice Programmers}.
\newblock \bibinfo{journal}{\emph{ACM Trans. Comput.-Hum. Interact.}}
  (\bibinfo{year}{2023}).
\newblock


\bibitem[Price et~al\mbox{.}(2016)]%
        {price2016generating}
\bibfield{author}{\bibinfo{person}{Thomas~W Price}, \bibinfo{person}{Yihuan
  Dong}, {and} \bibinfo{person}{Tiffany Barnes}.}
  \bibinfo{year}{2016}\natexlab{}.
\newblock \showarticletitle{Generating data-driven hints for open-ended
  programming.}
\newblock \bibinfo{journal}{\emph{Int. Educ. Data Mining Society}}
  (\bibinfo{year}{2016}).
\newblock


\bibitem[Price et~al\mbox{.}(2019)]%
        {price2019comparison}
\bibfield{author}{\bibinfo{person}{Thomas~W Price}, \bibinfo{person}{Yihuan
  Dong}, \bibinfo{person}{Rui Zhi}, \bibinfo{person}{Benjamin Paa{\ss}en},
  \bibinfo{person}{Nicholas Lytle}, \bibinfo{person}{Veronica Catet{\'e}},
  {and} \bibinfo{person}{Tiffany Barnes}.} \bibinfo{year}{2019}\natexlab{}.
\newblock \showarticletitle{A comparison of the quality of data-driven
  programming hint generation algorithms}.
\newblock \bibinfo{journal}{\emph{International Journal of Artificial
  Intelligence in Education}}  \bibinfo{volume}{29} (\bibinfo{year}{2019}),
  \bibinfo{pages}{368--395}.
\newblock


\bibitem[Price et~al\mbox{.}(2017)]%
        {price2017factors}
\bibfield{author}{\bibinfo{person}{Thomas~W Price}, \bibinfo{person}{Zhongxiu
  Liu}, \bibinfo{person}{Veronica Catet{\'e}}, {and} \bibinfo{person}{Tiffany
  Barnes}.} \bibinfo{year}{2017}\natexlab{}.
\newblock \showarticletitle{Factors influencing students' help-seeking behavior
  while programming with human and computer tutors}. In
  \bibinfo{booktitle}{\emph{Proc. of ICER}}. \bibinfo{pages}{127--135}.
\newblock


\bibitem[Rivers and Koedinger(2017)]%
        {rivers2017data}
\bibfield{author}{\bibinfo{person}{Kelly Rivers} {and}
  \bibinfo{person}{Kenneth~R Koedinger}.} \bibinfo{year}{2017}\natexlab{}.
\newblock \showarticletitle{Data-driven hint generation in vast solution
  spaces: a self-improving python programming tutor}.
\newblock \bibinfo{journal}{\emph{International Journal of Artificial
  Intelligence in Education}}  \bibinfo{volume}{27} (\bibinfo{year}{2017}),
  \bibinfo{pages}{37--64}.
\newblock


\bibitem[Sarsa et~al\mbox{.}(2022)]%
        {sarsa2022automatic}
\bibfield{author}{\bibinfo{person}{Sami Sarsa}, \bibinfo{person}{Paul Denny},
  \bibinfo{person}{Arto Hellas}, {and} \bibinfo{person}{Juho Leinonen}.}
  \bibinfo{year}{2022}\natexlab{}.
\newblock \showarticletitle{Automatic Generation of Programming Exercises and
  Code Explanations Using Large Language Models}. In
  \bibinfo{booktitle}{\emph{Proc. of ICER}}. \bibinfo{pages}{27--43}.
\newblock


\bibitem[Shute(2008)]%
        {shute2008focus}
\bibfield{author}{\bibinfo{person}{Valerie~J Shute}.}
  \bibinfo{year}{2008}\natexlab{}.
\newblock \showarticletitle{Focus on formative feedback}.
\newblock \bibinfo{journal}{\emph{Review of educational research}}
  \bibinfo{volume}{78}, \bibinfo{number}{1} (\bibinfo{year}{2008}),
  \bibinfo{pages}{153--189}.
\newblock


\bibitem[Vaithilingam et~al\mbox{.}(2022)]%
        {vaithilingam2022expectation}
\bibfield{author}{\bibinfo{person}{Priyan Vaithilingam},
  \bibinfo{person}{Tianyi Zhang}, {and} \bibinfo{person}{Elena~L Glassman}.}
  \bibinfo{year}{2022}\natexlab{}.
\newblock \showarticletitle{Expectation vs. experience: Evaluating the
  usability of code generation tools powered by large language models}. In
  \bibinfo{booktitle}{\emph{CHI conference extended abstracts}}.
  \bibinfo{pages}{1--7}.
\newblock


\bibitem[Zhang et~al\mbox{.}(2022)]%
        {zhang2022repairing}
\bibfield{author}{\bibinfo{person}{Jialu Zhang}, \bibinfo{person}{Jos{\'e}
  Cambronero}, \bibinfo{person}{Sumit Gulwani}, \bibinfo{person}{Vu Le},
  \bibinfo{person}{Ruzica Piskac}, \bibinfo{person}{Gustavo Soares}, {and}
  \bibinfo{person}{Gust Verbruggen}.} \bibinfo{year}{2022}\natexlab{}.
\newblock \showarticletitle{Repairing Bugs in Python Assignments Using Large
  Language Models}.
\newblock \bibinfo{journal}{\emph{arXiv preprint arXiv:2209.14876}}
  (\bibinfo{year}{2022}).
\newblock


\end{thebibliography}
